\documentclass[aps,draft,twocolumn]{revtex4}

\usepackage{color}

\input epsf

\begin{document}

\title{
Different types of the dimensional crossover in the quasi-one-dimensional
spinless fermion systems
}
\author{A.V. Rozhkov}

\affiliation{
Institute for Theoretical and Applied Electrodynamics RAS,
Moscow, ul. Izhorskaya 13, 125412, Russian Federation
}

\begin{abstract}
It is known that many-body correlations qualitatively modify the properties
of a one-dimensional metal. However, for a quasi-one-dimensional metal
these correlations are suppressed, at least partially. We study conditions
under which the one-dimensional effects significantly influence the
dimensional crossover of a quasi-one-dimensional metal. It is proved (i)
that even a system with very high anisotropy of the single-particle hopping
might behave on both sides of the crossover as an ordinary weakly non-ideal
Fermi gas. Further, (ii) to demonstrate well-developed signatures of
one-dimensional correlations the system must have extremely (exponentially)
high anisotropy. Between cases (i) and (ii) an intermediate regime lies: 
(iii) the one-dimensional phenomena affect the two-particle
susceptibilities, but do not reveal themselves in single-particle
quantities. Unlike the normal state properties, (iv) the
ordering transition is always very sensitive to the anisotropy: the mean
field theory quickly becomes invalid as the anisotropy increases. An
expression for the transition temperature is derived. The attributes (i-iv)
are used to classify the weakly interacting quasi-one-dimensional fermion
systems.
\end{abstract}

\maketitle
\hfill
\section{Introduction}

Several physical systems may be viewed as quasi-one-dimensional (Q1D)
fermionic liquids. These include materials with the Q1D anisotropy of the
electron hopping(e.g., Bechgaard salts, blue bronzes 
\cite{gruner}),
cold atoms in an anisotropic trap
\cite{cold_atoms2005},
and the artificially created atomic lattices 
\cite{zaizot_ref}.

The most universal feature of the Q1D systems is the dimensional crossover
(DC): at the temperatures exceeding some characteristic scale
$T_x$
a system behaves as an array of almost independent one-dimensional (1D)
units, while below
$T_x$
a genuine 3D behavior is recovered. The description of this crossover is of
fundamental importance for the theory of the systems in question.

Theoretically, the Q1D systems are frequently pictured as a lattice of 1D
chains, each chain is represented by a 1D Tomonaga-Luttinger model
\cite{boson},
and closely located chains are coupled by the weak transverse
single-electron hopping, or the weak interchain interaction, or both.  It
is often assumed that the DC occurs between the Tomonaga-Luttinger liquid
at high energy and the three-dimensional (3D) Fermi liquid at low energy.
Since the Tomonaga-Luttinger excitations are usually represented in terms
of the bosonic quantum numbers (however, see
Refs.~\cite{mattias_lieb1965,fermi_unit1,fermi_unit2,fermi_unit3,
imambekov_glazman_science2009}),
one has to describe how the 1D bosons cross over to the 3D fermions. This
is, of course, a difficult task.

A diverse set of the many-body tools has been used to study the DC.
Analytical renormalization group (RG) is applied in
Refs.~\cite{prigodin_firsov_a,bourbonnais_caron_rg_analyt,bourbonnais_caron}.
Numerical RG is employed in
Refs.~\cite{duprat2001,nickel2005,fuseya2007,fuseyaII2007}.
Modification of the dynamical mean field theory to the Q1D fermions is used
in
Ref.~\cite{dmft_q1d}.
Variational technique which explicitly construct both high-energy boson
excitations and low-energy fermion excitations is proposed in
Refs.~\cite{rozhkov_spinless_variational,rozhkov_variational}.
Different versions of the random phase approximations (RPA) are also used,
Refs.~\cite{aizawa,kuroki_rpa2004}.

However, the Tomonaga-Luttinger-liquid-based approaches to the crossover
may, in some situations, overcomplicate the theory. It is important to
realize that the DC, by itself, is not a many-body phenomenon. Instead, its
origin is purely kinematic: it occurs when the temperature becomes
comparable to the transverse electron hopping. As such, it occurs even for
systems with no interaction
\cite{free_el_chapter,free_el_arxiv}.
Thus, the presence of the crossover does not immediately imply that the
the high-energy phase is fundamentally different from the low-energy phase.

The free Q1D system is, of course, a trivial example. However, it may be
generalized to a less obvious case of the weakly interacting system.
Specifically, we will prove in this paper that in a broad parameter region
the Q1D fermions on both sides of the DC are closer to the weakly non-ideal
Fermi gas than to a collection of the Tomonaga-Luttinger liquids.  Further,
we demonstrate that, as parameters are varied, the DC itself experiences
several crossovers. It evolves from Fermi-liquid-to-Fermi-liquid type at
low anisotropy and interaction to Tomonaga-Luttinger-liquid-to-Fermi-liquid
at high anisotropy and interaction, with a more exotic possibility in
between. 

These different types of the DC may be characterized in terms of the
applicability of the low-order perturbation theory. The crossover between
the Tomonaga-Luttinger and the Fermi liquid may not be described by the
perturbation theory. One can deduce that from the fact that the
Tomonaga-Luttinger state is non-perturbative. As the anisotropy or
interaction decreases, the applicability of the perturbation theory
improves: below certain limit, the perturbation theory can be used for the
single-particle properties, but not for the two-particle properties. When
even the two-particle properties are within the range of the perturbation
theory, the Q1D fermions behave as a Fermi liquid both at high and low
energies. Note that such conductor may have very high anisotropy.

Finally, we investigate the applicability of the mean field theory for the
Q1D fermions. Apparently, if the anisotropy is large an ordering transition
is not of the mean field character. However, one expects that below a
certain point the mean field theory becomes accurate. It is surprising to
discover that even for a system whose normal state is well-described by the
perturbation theory the mean field theory may fail. We will prove that the
mean field theory works only if the hopping anisotropy is of the order of
unity
\cite{q1d_mf}. 

%
%
%

The paper is organized as follows. In 
Sec.~\ref{sect::model}
the model under study is described. We derive the condition which
guarantees the validity of the perturbation theory for the single-particle
properties in
Sec.~\ref{sect::1p}.
In
Sec.~\ref{sect::2p}
similar condition for the two-particle properties is established. In
Sec.~\ref{sect::mft}
the applicability of the mean field theory is discussed. The results of
these sections are used in
Sec.~\ref{sect::classify}
to introduce a classification of the anisotropic Fermi liquids.
The results are discussed in
Sec.~\ref{sect::disc}.
Sec.~\ref{sect::conclusions}
is reserved for the conclusions.


\section{Model}
\label{sect::model}
We study the following system of the interacting spinless fermions:
\begin{eqnarray}
\label{H}
&&H
= 
\sum_i \int_0^L dx {\cal H}_i^{\rm 1d} 
+ 
\sum_{i,j} \int_0^L dx {\cal H}_{ij}^\perp,
\\
&&{\cal H}_i^{\rm 1d}
=
{\rm i} v_{\rm F}
\left(
	\colon
	\psi^\dagger_{{\rm L}i}
	\nabla\psi^{\vphantom{\dagger}}_{{\rm L}i}
	\colon
	- 
	\colon
	\psi^\dagger_{{\rm R}i}
	\nabla\psi^{\vphantom{\dagger}}_{{\rm R}i}
	\colon
\right)
+
g \rho_{\rm R} \rho_{\rm L},
\\
&&{\cal H}_{ij}^\perp
=
- t(i-j)
\sum_{p={\rm L,R}}
	\psi^\dagger_{pi}\psi^{\vphantom{\dagger}}_{pj}
	+
	{\rm h.c.},
\label{perp}
\\
&&\rho_{pi}
=
\colon
\psi^\dagger_{pi}
\psi^{\vphantom{\dagger}}_{pi}
\colon,
\end{eqnarray}
where the fermionic field
$\psi_{pi}^\dagger$
creates a physical fermion with the chirality
$p={\rm L}$
or
$p={\rm R}$
on chain $i$. The chains are parallel to each other and form a 1D or 2D
square lattice in the directions transverse to the chains. Colons stand for
the normal ordering. The microscopic cutoff of the model is denoted by
$\Lambda$. The transverse tunneling amplitudes 
$t(i-j)$
depend on the distance 
$|i-j|$ 
between the chains. If
$t=0$
our Hamiltonian corresponds to a number of decoupled Tomonaga-Luttinger
systems. Below we will assume that 
$t(i-j)$ 
is non-zero for the nearest neighbors only. Generalization beyond this
assumption does not bring new features to the discussion.

The dimensionless interaction parameter is required to be small:
\begin{eqnarray}
\label{weak_inter}
\tilde g
=
\frac{
	g
     }
     {
	2 \pi v_{\rm F}
     }
\ll
1.
\end{eqnarray}
Here $g$ is the bare interaction strength, and
$v_{\rm F}$
is the Fermi velocity.

In principle, the transverse interactions can be also considered.
Sufficiently strong transverse interactions may trigger symmetry-breaking
phase transition at the temperature exceeding the single-particle
DC. We do not want to study this regime, and assume that
the transverse interactions are zero.

\section{Perturbation theory for single-particle properties}
\label{sect::1p}

\subsection{General remarks on the perturbation theory for the Q1D systems}
\label{subsect::remarks}

For a generic Fermi system the smallness of the interaction constant is a
sufficient condition for the applicability of the perturbation theory.
While the perturbative expansion contains divergent terms (e.g., the Cooper
diagram grows logarithmically for
$T \rightarrow 0$),
these divergences are well-understood, and the recipes for the perturbative
calculations of the physically relevant quantities are known.

Unfortunately, this program cannot be directly adopted for a Q1D Fermi
system: some perturbation theory terms, while small for a generic Fermi
liquid, in a Q1D case may be finite, but parametrically large. This
phenomenon occurs because the pure 1D fermion system has additional
divergent diagrams which are finite for the higher-dimensional Fermi
liquid. In Q1D system, the latter divergences are capped by arbitrary weak
transverse hopping, yet, the diagram values are affected by the proximity
to the divergences. Applying blindly the usual schemes in such a situation
may lead to significant qualitative errors in the estimation of the
effective parameters of the system. We will study below what limitations
should be placed on the microscopic constants of our model to guarantee
that these diagrams remain small, and the generic perturbation theory
procedures may be implemented.

In addition to the purely mathematical, formal side, the perturbation
theory applicability criteria carry important physical information about
the DC. When the system experience the crossover from the
Tomonaga-Luttinger liquid at high energy to the Fermi liquid at low energy,
the perturbation theory is useless due to non-perturbative nature of the
Tomonaga-Luttinger liquid. On the other hand, if the perturbation theory is
applicable, the system behaves as a Fermi liquid even above the crossover.
An accurate analysis reveals that a more exotic possibility is also
possible. With this considerations in mind we start our study of the
perturbation theory for the Q1D fermions.

\subsection{Diagram evaluation approach}
\label{subsect::diagram}

\begin{figure} [t]
\centering
\leavevmode
\epsfxsize=8.5cm
\epsfbox{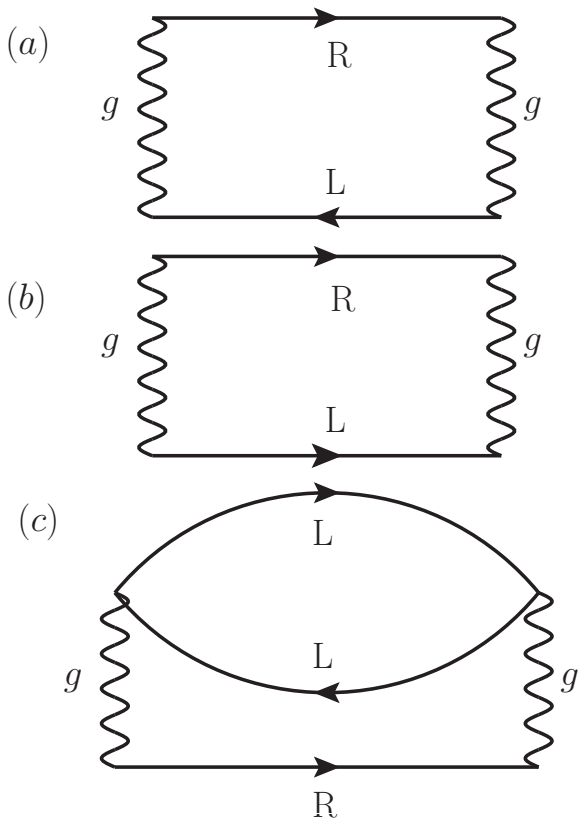}
\caption[]
{\label{fig::diags}
Three divergent diagrams of the 1D metal. Solid lines with arrow and `R/L'
chirality labels are fermion propagators. Wiggly lines represent
interaction $g$. Panel (a): the ``bubble" diagram corresponds to the
scattering in the charge-density-wave channel; panel (b): the Cooper
diagram describes the scattering in the superconducting channel; panel (c):
the single-particle self-energy diagram. Other divergent diagrams can be
obtained from (a-c) by inverting the chirality labels or directions of the
arrows on the fermion lines. 
}
\end{figure}

If we were to use the perturbation theory in order of
$\tilde g$
to study Hamiltonian (\ref{H}),
we would discover that, if
$t = 0$,
then several irreducible diagrams are divergent. Three of them are shown in
Fig.~\ref{fig::diags}.
Others can be constructed from these by inverting the chirality labels or
directions of the arrows on the fermion lines.

For example, at 
$T=0$
the self-energy correction [diagram (c) of 
Fig.~\ref{fig::diags}]
for the electrons of chirality $p$ is equal to
\begin{eqnarray}
\Sigma_p 
&=&
\frac{g^2}{16 \pi^2 v_{\rm F}^2}
(\omega - p v_{\rm F} k + i0 )
\\
\nonumber
&&\times
\ln \left[
		\frac{
			v_{\rm F}^2 k^2 - (\omega + i0)^2
		     }
		     {
			4 v_{\rm F}^2 \Lambda^2
		     }
    \right] + \ldots,
\end{eqnarray}
where the ellipsis stands for the non-singular terms, $k$ is the momentum
parallel to the chains.

This self-energy contributes to the renormalization of the quasiparticle
residue 
$Z_p = 1 - \delta Z_p$:
\begin{eqnarray}
\label{Z_1d}
\delta Z_p 
=
\frac{g^2}{16 \pi^2 v_{\rm F}^2}
\ln \left|
		\frac{
			4 v_{\rm F}^2\Lambda^2
		}
		{
			v_{\rm F}^2 k^2 - \omega^2
		}
    \right| + \ldots
\end{eqnarray}
The logarithmic divergence of 
$\delta Z_p$
on the mass surface implies the breakdown of the Landau theory of the Fermi
liquid for 1D Fermi systems.

For Q1D system 
Eq.~(\ref{Z_1d}) 
is not applicable near the mass surface. Indeed, the divergence of the
self-energy is purely 1D effect
\cite{maslov2005}.
Therefore, the growth of 
$\delta Z_p$
at small energy and momenta is cut off at the scale 
$\sim t$:
\begin{eqnarray}
\label{Z_q1d}
\delta Z_p ( \omega, k, {\bf k}_\perp)
<
\delta Z_p^{\rm max} 
\approx
\frac{g^2}{8 \pi^2 v_{\rm F}^2}
\ln \left(
		\frac{v_{\rm F} \Lambda}{t}
    \right).
\end{eqnarray}
A detailed derivation of this result is given in
Ref.~\onlinecite{rozhkov_spinless_variational}.

Equation~(\ref{Z_q1d})
can be used to determine the applicability limits of the perturbation
theory: the expansion in orders of 
$\tilde g$
is valid if 
\begin{eqnarray}
\label{1p_cond}
\delta Z_p^{\rm max}
\ll
1
\Leftrightarrow
t
\gg
t^{\rm 1P},
\\
t^{\rm 1P}
=
v_{\rm F} \Lambda
\exp
\left(
	-\frac{2}{\tilde g^2}
\right).
\end{eqnarray}
Thus, if the transverse tunneling exceeds the exponentially small value
$t^{\rm 1P}$,
the self-energy diagram is not only finite, but also small. Smallness of 
$\delta Z_p$
implies that the perturbatively defined fermionic quasiparticles are ``good"
excitations of our system both above and below the crossover.

Our calculations, however, do not evaluate higher-order contributions to
$Z$. They can be easily found with the help of a different approach, which
will be presented in the next subsection.

\subsection{Renormalization group argument}
\label{subsect::rg}

The applicability of the perturbation theory may be discussed using
different type of reasoning. Specifically, consider the
renormalization-group (RG) flow near the Tomonaga-Luttinger fixed point.
When the cutoff is reduced from $\Lambda$ to
$\tilde \Lambda$,
the effective value of $t$ becomes
\begin{eqnarray}
\label{eff_t}
\tilde t_{\tilde \Lambda} = t \left(
				\frac{\tilde \Lambda}{\Lambda}
		      \right)^\theta,
\end{eqnarray}
where 
$\theta \approx \tilde g^2/2$
is the anomalous dimension of the hopping operator. This RG scaling is
applicable until
$v_{\rm F} \tilde \Lambda
\propto
\tilde t_{\tilde \Lambda}$,
at which point
\begin{eqnarray}
\label{t_eff}
\tilde t = t \left(
			\frac{t}{v_{\rm F} \Lambda}
	     \right)^{\frac{\theta}{1-\theta}}
=
t \exp\left[
			{\frac{\theta}{1-\theta}}
			\log\left(
				\frac{t}{v_{\rm F} \Lambda}
			    \right)	
   \right].
\end{eqnarray}
This formula may be used to evaluate the DC scale:
$T_x \propto \tilde t$.
At energies below
$T_x$
one cannot view the system as 1D even approximately. Rather, it behaves as
the anisotropic multidimensional (2D or 3D) Fermi liquid with the effective
hopping 
$\tilde t$
and cutoff
$\tilde \Lambda$.

At small $\theta$ one can attempt to expand the exponent in
Eq.~(\ref{t_eff}):
\begin{eqnarray}
\label{eff_t_approx}
\tilde t
=
t \left[
	1
	+
	\left(
		{\frac{\theta}{1-\theta}}
	\right)
	\ln \left(
			\frac{t}{v_{\rm F} \Lambda}
	   \right)
	+
	\ldots
  \right].
\end{eqnarray}
When condition
Eq.~(\ref{1p_cond})
is satisfied, this expansion is valid, and the renormalization of the
transverse hopping is small. Otherwise, $t$ experiences strong
renormalization which cannot be captured by 
Eq.~(\ref{eff_t_approx}),
and full 
Eq.~(\ref{t_eff})
must be used.

To establish the connection with the discussion of the previous subsection,
observe that the effective hopping may be written as
\begin{eqnarray}
\tilde t = t Z_p = t + \delta Z_p t.
\end{eqnarray}
Therefore, the expression in the square brackets in 
Eq.~(\ref{eff_t_approx})
is nothing but the expansion of
$Z_p$
in orders of
$\theta=O(\tilde g^2)$
whose lowest order term is given by 
Eq.~(\ref{Z_q1d}).

The main advantage of the presented argument is that it automatically
accounts for the higher-order contributions: one can keep as many terms in
the expansion
Eq.~(\ref{eff_t_approx})
as needed. Furthermore, this approach makes the statement of this section
almost arithmetical: in order to measure reliably the exponent $\alpha$ of
a power-law function
$f(x)=x^\alpha$
one must sample the function $f$ over an exponentially large range of $x$.
For example, the transverse conductivity 
$\sigma_y$
at
$v_{\rm F} \Lambda > T > T_x$
demonstrates the non-universal power-law behavior
\cite{lopatin2001}:
\begin{eqnarray}
\label{eq::sigma}
\sigma_y
\propto
\left(
	\frac{T}{v_{\rm F} \Lambda}
\right)^{-1+2 \theta}.
\end{eqnarray} 
However, this non-universality may be detected only if the ratio
$(T_x/v_{\rm F} \Lambda)$
is exponentially large. Otherwise, 
Eq.~(\ref{eq::sigma})
is indistinguishable from 
(see Eq.~(71) of Ref.~\cite{lopatin2001})
\begin{eqnarray}
\label{univ_cond}
\sigma_y \propto 1/(g^2 T)
\end{eqnarray} 
with weak corrections. 
Equation~(\ref{univ_cond})
contains only universal exponents. Thus, on experiment the universal
transverse transport indicates the validity of 
Eq.~(\ref{1p_cond})
and the applicability of the perturbation theory for the single-particle
propagator.

\section{Perturbation theory for two-particle properties}
\label{sect::2p}

\subsection{Diagram evaluation approach}
\label{subsect::diagram_2p}

The one-dimensional effects affect the two-particle properties as well.
Consider the diagrams (a) and (b) in
Fig.~\ref{fig::diags}.
They contribute to the renormalization of the effective interaction.
Diagram (a) represents the scattering of a particle-hole pair, while
diagram (b) corresponds to the Cooper pair scattering.

In a generic Fermi liquid diagram (b) has logarithmic divergence when the
total momentum of the Cooper pair is zero. For attractive interaction this
divergence leads to the Cooper instability. As for diagram (a), it diverges
at the nesting vector, provided that the Fermi surface nests well. This
diagram is responsible for the density wave instability. For a generic
Fermi liquid the channels are said to be decoupled in the sense that the
Cooper pair diagram is finite and small in the particle-hole channel, while
the particle-hole diagram is finite and small for the Cooper pair
scattering.

In 1D systems, however, the channels are coupled: the particle-hole
contribution to the Cooper pair scattering is divergent; moreover, the
strength of this divergence is equal in magnitude and opposite in sign to
the divergence of the Cooper diagram (see Eqs.~(1.45) of 
Ref.~\cite{book_giamarchi}).
The same is true about the contribution of the Cooper diagram to the
particle-hole channel. This cancellation is a unique 1D feature responsible
for the coupling $g$ being exactly marginal.

In Q1D system the contribution of the Cooper diagram to the particle-hole
channel and vice versa are finite, but not necessarily small. Let us, for
definiteness, consider the particle-hole channel. For the sake of
simplicity, assume that the Q1D Fermi surface nests perfectly at the
nesting vector
${\bf Q}$.
The formal expression for the effective coupling at 
${\bf Q}$,
to the first order in 
$\tilde g$,
is:
\begin{eqnarray}
\label{g_corr}
g_{\rm eff}
\approx
g\left[
	1
	+
	\frac{g}{2\pi v_{\rm F}} \ln \left(
						\frac{v_{\rm F} \Lambda}{T}
					\right)
	-
	\frac{g}{2\pi v_{\rm F}} \ln \left(
						\frac{v_{\rm F} \Lambda}{t}
					\right)
\right].
\end{eqnarray}
The first term here is the bare coupling, the second term corresponds to
the ``bubble" diagram correction, and the third term is the contribution of
the Cooper diagram for
$T \ll t$.
Both corrections are small provided that
\begin{eqnarray}
\label{2p_cond}
t \gg t^{\rm 2P}, 
\\
\label{2p_T_cond}
T \gg t^{\rm 2P},
\\
\label{t_2p}
t^{\rm 2P} = v_{\rm F} \Lambda \exp \left(
						-\frac{2\pi v_{\rm F}}{g}
				    \right)
=
v_{\rm F} \Lambda \exp \left(
				-\frac{1}{\tilde g}
  		       \right).
\end{eqnarray}
If we were to consider the effective interaction in the Cooper channel, we
would, going through the same steps, obtain the same result.

Eqs.~(\ref{2p_cond}) and~(\ref{2p_T_cond})
define the parameter region in which the susceptibility may be calculated
perturbatively. 
Equation~(\ref{2p_cond})
ensures the destruction of the non-perturbative 1D effects.
Equation~(\ref{2p_T_cond})
must be enforced for a reason which has nothing to do with 1D phenomena:
below
$t^{\rm 2P}$
the non-perturbative physics of the approaching phase transition starts to
affect the susceptibility.

\subsection{Renormalization group argument}
\label{subsect::rg_2p}

It is instructive to rederive
Eq.~(\ref{2p_cond})
in a fashion similar to the one presented in
subsection~\ref{subsect::rg}.
To this end, consider the charge-density wave (CDW) susceptibility 
$\chi_{\rm CDW}$
for 
$T > T_x$ (see, e.g., Eq.~(1.68) of
Ref.~\cite{book_giamarchi}):
\begin{eqnarray}
\label{2p_scaling}
\chi_{\rm CDW} (T)
=
\frac{1}{2 \pi v_{\rm F} \tilde g}
\left[
	1
	-
\left(
	\frac{
		v_{\rm F} \Lambda
	     }
	     {
		T
	     }
\right)^{2 \tilde g}
\right] + \ldots,
\end{eqnarray}
where the ellipsis stands for non-singular contributions to the
susceptibility. Thus, the expansion in powers of $g$
\begin{eqnarray}
\chi_{\rm CDW}
=
-
\frac{1}{\pi v_{\rm F}}
\ln 
\left(
	\frac{
		v_{\rm F} \Lambda
	 }
	 {
		T
	 }
\right)
-
\frac{g}{2\pi^2 v_{\rm F}^2} 
\log^2 \left(
		\frac{v_{\rm F} \Lambda}{T} 
	\right)
+ 
\ldots,
\end{eqnarray}
is valid at
$T > T_x \sim t$,
provided that
Eq.~(\ref{2p_cond})
is fulfilled. 
	
Therefore, we conclude that, if 
Eq.~(\ref{2p_cond})
holds, the perturbation theory for two-particle quantities is applicable to
any order in $g$, and the crossover between the Tomonaga-Luttinger liquid
scaling
Eq.~(\ref{2p_scaling})
above $T_x$ and the Fermi liquid behavior below 
$T_x$
cannot be observed. To detect the high-energy scaling we must work with
with a sufficiently anisotropic, or sufficiently non-ideal system for which
Eq.~(\ref{2p_cond})
is violated.

\section{Applicability of the mean field theory}
\label{sect::mft}

The mean field theory is a valuable tool to study the phase diagrams of
interacting systems. Both the mean field theory (e.g.,
Refs.~\onlinecite{parish2007,maki2008,lebed2011},
chapter~4.4 of
Ref.~\onlinecite{organic_superconductors},
chapter~3 of
Ref.~\onlinecite{gruner})
and the closely-related RPA (e.g.,
Ref.~\onlinecite{kuroki_rpa2004})
have been used in the context of the Q1D fermions. Thus, applicability of
the mean field theory is an important issue for a theory of the Q1D Fermi
systems. 

As we have seen above, the perturbation theory may work well for the Fermi
liquids with high anisotropy. The mean field theory, however, is more
fragile: we will show that it can be applied only to a system with the
anisotropy of the order of unity. In addition, we will derive an expression
for the transition temperature which is valid if
Eq.~(\ref{2p_cond}) 
is true.

To notice the difficulty facing the mean field theory let us make the
following heuristic observation. For an anisotropic system two different
formulas for the transition scale can be constructed. The first one is
$t^{\rm 2P}$
(for example, the expression of this type is given by Eq.~(4.41) of
Ref.~\cite{organic_superconductors})
and the second one is
\begin{eqnarray}
\label{T_mf}
T_{\rm CDW}
=
t \exp \left(
		-\frac{2\pi v_{\rm F}}{g}
       \right)
=
\left(
	\frac{t}{v_{\rm F} \Lambda} 
\right)
t^{\rm 2P}.
\end{eqnarray} 
When
$t/v_{\rm F} \Lambda \sim 1$
the two answers coincide up to a factor of the order of unity, which is a
typical accuracy of the mean field theory (see, e.g., 
Ref.~\cite{kos_millis_larkin}).
However, if 
$t/v_{\rm F} \Lambda \ll 1$,
then 
$T_{\rm CDW} \ll t^{\rm 2P}$,
and one has to decide which of the two is valid. As it turns out,
Eq.~(\ref{T_mf})
is the right answer.

The discussion of the previous paragraph may be cast in a more formal
fashion. When 
inequality~(\ref{2p_cond})
holds the third term in
Eq.~(\ref{g_corr})
is much smaller than unity and, superficially, can be neglected.

Once it is neglected the geometrical progression of the divergent ``bubble"
diagrams contributing to 
$g_{\rm eff}$
must be summed. If 
$g>0$
the coupling 
$g_{\rm eff}$
found in such a manner diverges at
$T=t^{\rm 2P}$,
signaling the transition into the CDW state.

This argumentation is invalid, however. The omission of the third term of
Eq.~(\ref{g_corr}) 
on the grounds of its smallness is the offending step. Since the mean field
transition temperature is a non-analytical function of the interaction,
even a small correction to the latter may lead to large variation of the
former.

This property is not unique to the Q1D metallic system. The transition
temperature of the Kohn-Luttinger superconductor shows similar sensitivity
to the higher-order corrections to the coupling constant (see, e.g.,
Eq.~(31) of
Ref.~\cite{efremov_kohn_lutt}).

The correct way to address this issue for a Q1D metal is discussed in
Ref.~\cite{q1d_mf}.
Namely, one must perform the RG transformation until the crossover scale 
$T_x$
is reached. No abnormality pertinent to 1D physics is present below 
$T_x$.
When the effective Hamiltonian at the crossover scale is found, the mean
field calculations can be safely applied to it, and 
Eq.~(\ref{T_mf})
is recovered.

Alternatively, one can resort to the approach used in the theory of the
Kohn-Luttinger superconductivity 
\cite{kohn_lutt,kagan_chubukov}:
before summing the infinite series of divergent diagrams (Cooper diagram in
the case of the superconductivity, the ``bubble" diagram in our case), one
must account for renormalization of the coupling constant, which is
perturbatively dressed by a set of non-divergent diagrams. Thus, the
effective CDW coupling, which includes the Cooper diagram contribution, is
\begin{eqnarray}
\label{g_eff_pp}
g_{\rm eff}^{\rm CDW} 
\approx
g\left[
	1
	-
	\frac{g}{2\pi v_{\rm F}} \ln \left(
						\frac{v_{\rm F} \Lambda}{t}
				    \right)
\right].
\end{eqnarray}
Performing the summation of the ``bubble" series we obtain for the
transition temperature:
\begin{eqnarray}
\label{T_c_mf_deriv}
T_{\rm CDW}
=
v_{\rm F} \Lambda
\exp\left(
		-\frac{2 \pi v_{\rm F} }{g_{\rm eff}^{\rm CDW}}
    \right),{\rm \ where}
\\
\frac{1}{g_{\rm eff}^{\rm CDW} }
\approx
\frac{1}{g}
+
\frac{1}{2\pi v_{\rm F}} \ln \left(
					\frac{v_{\rm F} \Lambda}{t}
				\right),
\end{eqnarray} 
from which 
$T_{\rm CDW}$
is recovered.

Our discussion demonstrates that the anisotropy strongly affects the
transition temperature renormalizing it down from the prediction of the
mean field theory. We also learned that, when studying the phase diagram of
a Q1D Fermi system, a careful analysis of the theory's ``diagrammatic
content" is necessary. An indiscriminate use of a technique, performing
well for a generic Fermi liquid, can lead to a qualitative error for the
Q1D system.

\section{Four types of the anisotropic Fermi systems}
\label{sect::classify}

In the previous section we defined two energy scales,
$t^{\rm 1P}$
and
$t^{\rm 2P}$.
It is trivial to prove that
$
t^{\rm 1P}
\ll
t^{\rm 2P}
\ll
v_{\rm F} \Lambda.
$
Depending on how $t$ compares against these scales we may define four types
of the weakly-interacting anisotropic Fermi systems, see
Table~\ref{tab::types}.
\begin{table}
\begin{tabular}{||c|c|c||}
\hline\hline
Type 	& Hopping $t$	& Properties 	\cr
\hline\hline
I	& $t \ll t^{\rm 1P} \ll t^{\rm 2P}$
			& crossover: Tomonaga-Luttinger \cr
	&		& to Fermi liquid;\cr
	&		& small quasiparticle residue $Z_p \ll 1$; \cr
	&		& mean field theory is not applicable\cr
\hline
II	& $t^{\rm 1P}\ll t \ll t^{\rm 2P}$
			& crossover: shows 1D correlations \cr
	&		& in the susceptibilities only; \cr
	&		& quasiparticle residue $Z_p \sim 1$; \cr
	&		& mean field theory is not applicable\cr
\hline
III	& $t^{\rm 2P} \ll t \ll v_{\rm F} \Lambda$
			& crossover: Fermi to Fermi liquid;\cr
	&		& quasiparticle residue $Z_p \sim 1$; \cr
	&		& mean field theory is not applicable\cr
\hline
IV	& $t \propto v_{\rm F} \Lambda $
			& crossover: poorly developed;\cr
	&		& quasiparticle residue $Z_p \sim 1$; \cr
	&		& mean field theory is applicable\cr
\hline\hline
\end{tabular}
\caption{
Four types of the anisotropic Fermi liquids. For extremely anisotropic
type~I Fermi liquid both single-particle and two-particle quantities
experience strong renormalization due to 1D many-body effects
(quasiparticle residue is small; at high energies both the transverse
conductivity and the susceptibilities show power-law behavior with
non-universal exponents). Less anisotropic type~II system shows no 1D
effects in its single-particle properties. Consequently, its quasiparticle
residue is close to unity. However, the corrections to the susceptibilities
introduced by 1D effects are substantial, and cannot be accounted for by
the perturbation theory. Type~III Fermi liquid may be accurately described
by the finite-order perturbation theory. Yet, the mean field theory fails.
Finally, both the perturbation theory and the mean field theory works for a
type~IV system, whose anisotropy is of the order of unity. Due to poor
separation of $t$ and
$v_{\rm F} \Lambda$
the dimensional crossover is not well-defined.
}
\label{tab::types}
\end{table}

When the system parameters are such that
Eq.~(\ref{1p_cond})
is violated, we have the Fermi liquid of type~I. For such a system both
single-particle and two-particle quantities experience strong
renormalization due to 1D many-body effects. As follows from
Eq.~(\ref{1p_cond}),
the quasiparticle residue is small. The DC occurs between the high energy
phase of the Tomonaga-Luttinger liquid and the low energy phase of the 3D
anisotropic Fermi liquid. The mean field theory is not applicable.

The less anisotropic type~II system
[Eq.~(\ref{1p_cond})
is valid,
Eq.~(\ref{2p_cond})
is not] is very peculiar. It shows no 1D effects in its single-particle
properties. Consequently, its quasiparticle residue is close to unity. At
the same time, the susceptibilities demonstrate power-law scaling with a
non-universal exponent above the DC. Thus, the high-energy phase does not
show full phenomenology of the Tomonaga-Luttinger liquid. The mean field
theory is not applicable.

A type~III Fermi liquid 
[Eq.~(\ref{2p_cond}) 
is valid, but
$t \ll v_{\rm F} \Lambda$]
may be accurately described by the finite-order perturbation theory. Note
that such Fermi liquid is strongly anisotropic, that is, the strong
anisotropy alone is not sufficient for the system to show the 1D many-body
effects. However, for type~III system, the mean field theory is not
applicable.

Finally, type~IV Fermi liquid 
($t \lesssim v_{\rm F} \Lambda$)
can be described by both the perturbation theory and the mean field theory.
However, due to poor separation of the transverse and longitudinal kinetic
energy scales the DC is not well-defined for this class of systems.

\section{Discussion}
\label{sect::disc}

Of the four types of the Q1D systems type~I is the most difficult to
observe in the weak coupling regime. For example, if
$\tilde g = 0.3$
the anisotropy ratio must be very high:
\begin{eqnarray}
\frac{v_{\rm F} \Lambda}{t} > 2 \cdot 10^{10}.
\label{typeI}
\end{eqnarray}
At smaller $\tilde g$ it must be even higher. 

The latter conclusion does not contradict the fact that the Bechgaard
salts, whose anisotropy ratio is 10, demonstrate the type~I phenomenology 
(the transverse resistivity shows power-law behavior with a non-universal
exponent
\cite{dressel}).
One must remember that these compounds are in the intermediate, not weak,
coupling range, that is,
inequality~(\ref{weak_inter})
is violated. In the intermediate coupling regime the transverse transport
is non-universal if [see
Eq.~(\ref{eff_t_approx})]
\begin{eqnarray} 
\label{1p_cond_inter}
\left(
	{\frac{\theta}{1-\theta}}
\right)
\ln \left(
		\frac{v_{\rm F} \Lambda}{t}
   \right) 
> 1.
\end{eqnarray}
Here, instead of expanding $\theta$ in powers of
$\tilde g$
which would lead us to 
Eq.~(\ref{1p_cond}),
we kept the full functional dependence as it appears in
Eq.~(\ref{eff_t_approx}).
When
$\tilde g = O(1)$
Eq.~(\ref{1p_cond_inter})
is more accurate than
Eq.~(\ref{1p_cond}).
For the anisotropy of 10,
Eq.~(\ref{1p_cond_inter})
is valid if
$\tilde g$
exceeds 0.64, or, equivalently, the Tomonaga-Luttinger parameter
${\cal K}$
is smaller than 0.47. This conclusion is consistent with the estimate 
${\cal K} \approx 0.23$
for (TMTSF)$_2$PF$_6$
\cite{dressel}.

The requirement for type~II is far less stringent than 
Eq.~(\ref{typeI}):
anisotropy must exceed the following values
\begin{eqnarray} 
\frac{v_{\rm F} \Lambda}{t^{\rm 2P}} \approx 30
{\rm \ for \ \ }
\tilde g = 0.3,
\\
\frac{v_{\rm F} \Lambda}{t^{\rm 2P}} \approx 150
{\rm \ for \ \ }
\tilde g = 0.2.
\end{eqnarray} 
Thus, it is likely that the crossovers between type II, III, and IV systems
may be realized experimentally.

To observe this sequence of the crossovers the Bechgaard salts are not
suitable. Indeed, for a Q1D conductor with the interaction of intermediate
strength 
\begin{eqnarray}
t \sim t^{\rm 1P} \sim t^{\rm 2P}.
\end{eqnarray}
Consequently, it is difficult to resolve the different liquid types. 

More promising for our purposes are the cold atoms in the Q1D optical trap.
The experimental implementation of this system has been reported in
Ref.~\cite{cold_atoms2005}.
The advantage of the cold atoms setup is its tunability: for example, the
interaction between the atoms can be smoothly changed from attraction to
repulsion. This makes the trapped atoms an appealing alternative to the
solid state implementations of the Q1D fermions.

The artificially created atomic lattices
\cite{zaizot_ref}
is yet another interesting Q1D Fermi system. However, this research is at
the beginning stage yet.

\section{Conclusions}
\label{sect::conclusions}

We demonstrated that in the weakly non-ideal Q1D Fermi liquids, depending
on the interaction and the anisotropy, the dimensional crossover occurs in
one of the four types. When the anisotropy is extremely strong, the system
shows clear crossover from the Tomonaga-Luttinger to the higher-dimensional
Fermi liquid behavior. At smaller anisotropy the Tomonaga-Luttinger physics
may be observed only in the two-particle properties (e.g.,
susceptibilities). When the anisotropy decreases even further the
Tomonaga-Luttinger features cannot be noticed. In the latter case the
crossover is similar to the crossover in the anisotropic free fermion
system. It was also proved that the anisotropy must be exponentially large
in order to observe at least some of the Tomonaga-Luttinger many-body
effects. 

It is shown that for the mean field theory to be valid low anisotropy is
required. More broadly, our discussion demonstrated the need for careful
analysis of the diagrammatic structure of a theoretical technique used for
study of the phase transition: we have seen that even small diagrammatic
contributions to the susceptibility may drastically affect the calculated
value of the transition temperature.

The cold atoms in the Q1D optical trap is the likely candidate where the
different types of the crossover may be observed.

\section{Acknowledgements}

The author is partially supported by RFBR grants 11-02-00708-a,
10-02-92600-KO-a, and 09-02-00248-a.



%
%
%
%
%

\bibliographystyle{apsrev_no_issn_url.bst}
\bibliography{crossover}

\end{document}